\documentclass[aps,prl,floatfix,showpacs,twocolumn,tightenlines,superscriptaddress]{revtex4}

\usepackage{amsmath}
\usepackage{graphicx}
\usepackage{epsfig}

\begin{document}

\title{Universal quantum computation on the power of quantum non-demolition measurements}
\author{Kae Nemoto}\email{nemoto@nii.ac.jp}
\affiliation{National Institute of Informatics, 2-1-2 Hitotsubashi, Chiyoda-ku, Tokyo 101-8430, Japan}

\author{W. J. Munro}\email{bill.munro@hp.com}
\affiliation{Hewlett-Packard Laboratories, Filton Road, Stoke Gifford,
Bristol BS34 8QZ, United Kingdom}
\affiliation{National Institute of Informatics, 2-1-2 Hitotsubashi, Chiyoda-ku, Tokyo 101-8430, Japan}

\date{\today}

\begin{abstract}

In this letter we investigate the linear and nonlinear models of optical quantum 
computation and discuss their scalability and efficiency. We show how there are 
significantly different scaling properties in single photon computation when weak 
cross-Kerr nonlinearities are allowed to supplement the usual linear optical set. 
In particular we show how quantum non-demolition measurements are an efficient resource 
for universal quantum computation.

\end{abstract}


\maketitle

In classical computation, we can always decompose a circuit by a complete set of
gates, such as the AND, OR or NOT gates. These decompositions can be similarly 
applied to quantum computation and several sets of gates are known to be able 
to construct an arbitrary computational circuit to an arbitrary precision.  
These sets are called {\it a universal set of gates} and in principle, 
a universal set of gates with initializable qubits and projective measurement 
onto the computational basis states gives one the power to perform 
{\it universal quantum computation}.  This representation of the {\it universality 
of quantum computation} is gate-based in analogy to classical computation. 
A typical universal set of quantum gates could be arbitrary single qubit rotations 
and the CNOT gate (a two qubit gate)\cite{barenco95}. With these gates, the resources required for 
universal quantum computation are then initializable qubits, arbitrary single qubit 
rotations, the CNOT gate, projective measurements in the computational basis, and
classical feed-forward.

These universal sets of gates gives one a partial understanding of how quantum 
computers may differ from their classical counterparts. As the Gottesman-Knill 
theorem states\cite{gk99,gk99b,got04}, a computational circuit which consists of a set of gates 
including CNOT gate and some single-qubit rotations (the Hadamard and Pauli gates) 
can be efficiently simulated on a classical computer.  The gate set in the
Gottesman-Knill theorem is not universal as a gate set, however it is known that 
an additional operation such as the $\pi/8$-gate makes the gate set 
universal\cite{boykin00}. 
Another typical example of a universal gate sets is the pair of the Toffoli
Hadamard and $\pi/4$-gate \cite{kitaev97}, which is more computer-scientifically 
interesting as Toffoli gate 
is classically universal by itself.
Suppose that quantum model of computation is strictly more powerful than the classical one, 
then in general a quantum computation cannot be simulated efficiently on a classical computer. 
In fact, so far all the sets of gates which can be simulated on a 
classical computer are not universal. In this sense, universality 
can be used as a simple but useful criterion to distinguish quantum computational
circuits from classical computational ones.

Implementing a universal set of gates is certainly a way to achieve universal
quantum computation, however in real quantum mechanical systems, it is generally 
difficult to implement such a universal gate set. Generally speaking, systems 
with small decoherence have difficulties with controlled operations between two 
(or more) qubits and where such controlled operation are natural, it is often 
difficult to maintain quantum coherence in the system and to ensure the 
accessibility to each individual qubit. Given the fact that universal quantum 
computation requires both good quantum coherence and entanglement in the 
computational system, this trade-off relation between quantum coherence and
multi-qubit interaction seems to be an obstacle to universal quantum 
computation, when  one considers the requirement for universal gate sets. However,
implementation of universal gate sets is not the only way to achieve universal
quantum computation.  In fact, some implementations of non-universal gate sets 
can achieve universal quantum computation in a scalable manner. Such 
implementations are not allowed to access to the entire Hilbert space, yet is 
possible for these gates to simulate universal quantum computation on a lower 
dimensional subspace of a larger Hilbert space. The controlled-rotation gate allowing the qubit space to only
be in real coefficient states, which is a rebit subspace, is such an 
example\cite{rudolph02}. 

As long as we aim to achieve universal quantum computation, there is an 
alternative approach to it. Universal quantum computation can be achieved by 
measurement alone or are based on measurements\cite{Nielsen03,Leung04,Briegel01}. 
In classical computation, 
measurement schemes are trivial, and hence classically universal computation 
may be considered with gate sets alone, however in quantum computation this is 
not the case. Measurement alone can be as powerful as a universal gate set.  
There has been a number of measurement-based universal quantum computation 
schemes considered. The universal quantum computation can be broadly classified 
in three categories:
\begin{itemize}
\item{Universal sets:} A universal set can construct an
arbitrary circuit with an arbitrary precision.
\item{Universal computational set:} a universal computational set is not
universal by the definition of universality by not being able to construct an
arbitrary operation, however can simulate universal quantum 
computation in a larger Hilbert space with polynomial extra resource. This class of
sets does not allow the initial state to go to a certain subspace in the enitre 
Hilbert space. An example of this class is the rebit compuation in the previous 
paragraph.
\item{Non-universal sets:} a non-universal set can be efficiently simulated on
a classical computer and cannot simulate universal computation with polynomial extra
resource. It can construct universal computation only with additional measurement
strategies.
\end{itemize}

Now because of the power of measurement in quantum computation, it is obvious that 
it does not much make sense to evaluate quantum computation schemes solely either on 
gate sets or on measurement schemes. This aspect of quantum computation 
can be exploited to construct a circuit bypassing the difficulties on some 
particular operations, however at the same time it adds an extra complication into 
our criteria for universal quantum computation. The aim of this paper, instead of
using the conventional gate sets, is to introduce essential physical elements which inherit
fundamental physical properties required in universal quantum computation, which we
call physical primitives. The physical primitives are hence 
free from the concept of gates and measurements, and 
it is an advantage of this evaluation that we can specify the requirements of  the
physical properties for universal quantum computation.

There are typically three types of encoding we can consider in quantum information
processing. The most well investigated ones are the qubit and qudit encodings. 
These are discrete in nature and  can be mapped to each other relatively easily. 
The last qunat computation\cite{Lloyd99} can be quite different from qubit and qudit computations 
as it is based on continuous variable encoding rather than discrete bits or dits. 
As we aim to obtain a physical criteria to achieve universal quantum computation,
these differences should not be a matter for the criteria. It is necessary to merge
these different coding schemes in terms of requirements.  To do this, we first 
discuss optical qubit computation and then proceed to qunat computation. 

In this paper, we set optics as our primary physical system as the criteria make more sense
with a certain physical implementation, though the criteria are general. 
The typical universal set of gates for optical qubit-computation are single-qubit 
rotations and the two-qubit CNOT gate\cite{barenco95,KLM}.  There are obviously other combinations of 
fundamental logic gates for universality, however as we are focusing on the 
physical properties required to satisfy universality the combination of fundamental 
gates is not practically important. So we conveniently choose the universal set to
check universality of our new criteria. In optics (and especially with a 
polarization encoded qubit), single-qubit rotations are rather easy to and 
efficient to implement, but on the other hand two-qubit operations such as CNOT 
gate are hard to perform. We can set the typical computational requirements 
for single photon universal quantum computation as:
on demand single photon sources, arbitrary single-qubit rotations, a two qubit gate 
such as the CNOT gate, single photon counting and classical feed-forward.
Now, for this single photon based computation any two-qubit operation requires an 
optical nonlinearity, and hence the general difficulties in optical implementation 
arise from the lack of materials with an intrinsic optical nonlinearity. Entangling 
gates such as the CNOT are essential to perform the universal quantum computation 
and hence we require a mechanism to entangle the optical qubits. One well-known 
measurement-based scheme has been proposed by Knill, Laflamme and Milburn (KLM) who showed that 
a non-deterministic CNOT gate can be constructed using only single photon sources, linear optical elements 
and single photon number resolving detectors\cite{KLM,Pittman01,Ralph02,Knill02,Scheel03}. This probabilistic but 
heralded gate can then be teleported into the main stream quantum circuit enabling 
scalable computation\cite{KLM,Gottesman99,Bartlett03}, and hence the whole computation reminds scalable. 
Thus for this approach the physical devices for universal quantum computation are 
\vspace{-2mm}
\begin{itemize}
\item On demand single photon sources, \vspace{-2mm}
\item Linear optical elements,\vspace{-2mm}
\item Single photon counting, \vspace{-2mm}
\item Classical feed-forward. \vspace{-2mm}
\end{itemize}

Now, continuous variable (qunat-based) quantum computation\cite{Lloyd99} appears and looks rather 
different to this. First, the generalization of Gottesman-Knill theorem\cite{Bartlett02} states that 
a computational circuit staring from a computational basis state and using only 
operations with linear or quadratic  Hamiltonians\cite{footnote}, 
and finally measuring onto the computational basis states 
can be efficiently simulated on a classical computer. What is interesting here is that these quadratic Hamiltonians include entangling gates such as the SUM gate  (similar to the qubit CNOT gate). To make this gate/operation set universal, a 
third or higher order nonlinear Hamiltonian is required (the self-Kerr or cross-Kerr 
nonlinearities are such examples). This means the typical 
resources for universal continuous variable quantum computation are 
\vspace{-2mm}
\begin{itemize}\label{cv}
\item A coherent state source, \vspace{-2mm}
\item Linear and Quadratic Hamiltonian gates,\vspace{-2mm}
\item Homodyne measurement, \vspace{-2mm} 
\item classical feed-forward, \vspace{-2mm}
\item A third order or higher optical nonlinearity. \vspace{-2mm}
\end{itemize}
It is now obvious there is a difference between the qubit-based computation and qunat-based
computation in the way that qunat-entangling gate such as the SUM gate can be
implemented by linear elements, but by contrast the qubit CNOT gate is fundamentally 
nonlinear and hence in linear optics it has to be non-deterministically
constructed.  Universal qunat computation does require nonlinear elements/gates 
(though the form of nonlinearity is arbitrary) while linear optics quantum computation 
seems to remove the fundamental need for a nonlinear gate. This apparently suggests 
that there is also difference in physical resources between these two platforms. 
However, it is known that optical nonlinear elements/gates can be replaced by ideal 
photon counting in universal qunat computation\cite{Lloyd99}.  It is thus not clear whether or not 
these two sets of requirements can be merged to a common set of physical
requirements, i.e. physical primitives or these need to be different.  
To answer this question, we need to decompose these 
elements much further into physical primitives, focusing on the nonlinear optical 
elements. For instance, the equivalent effect of ideal photon counting to optical 
nonlinear gates may suggest that there is a sufficient amount of nonlinearity 
hiding in ideal photon 
counting, and probably also in single photon sources.  The ambiguities here may 
arise from the fact that single photon sources and detectors are not physically 
trivial elements.  It is thus necessary to examine the procedure for the 
generation of the single photons and their detection and identify the potential 
nonlinear elements embedded in them.

To achieve this, we will require all computation to begin with a coherent state and 
end with projective measurement on to a quadrature amplitude. These are all linear 
optical elements and operations and as such these is no hidden nonlinearity in the preparation procedure 
and the detection devices.  With these optically linear elements, this is a 
natural setting for qunat computation, however for qubit computation we need to 
have a mechanism to generate our initialized qubits and detect them. There are a number of ways 
this can be achieved and here we will discuss one of the theoretically simpler approaches. It is 
known that quantum non-demolition (QND) measurements can be used to generate 
and detect photon-number states\cite{Milburn84,Yamamoto87,Munro03}. The QND measurement 
can be constructed from an optical nonlinearity (such as a cross-Kerr nonlinearity) 
and a coherent state probe field and homodyne detection. The efficiency of the 
QND measurement as detector for qubit computation can be improved 
by either increasing the effective size of the optical nonlinearity or the amplitude 
of the coherent state $|\alpha\rangle$. 
The QND measurement requires a cross Kerr nonlinearity, and hence the requirements for 
qunats are applicable to qubit computation.  Therefore, on these grounds, in terms 
of universality of quantum computation, qubit-based computation and qunat-based 
computation are unified in one set of conditions. Linear optical elements and cross 
Kerr nonlinear coupling are physical primitives for universal quantum computation 
no matter the coding you have in in mind. 

These physical primitives are deterministic and hence might give a 
different scalability to the general linear optical quantum computation schemes.  
In linear optical quantum computation, it is usual that entangling gates have a 
very complicated structure. These gates are structured based on linear optics 
elements and measurement, hence in other words, linear optical quantum computation 
is measurement assisted computation without a universal set of gates. Only when the 
right measurement signature is obtained do the linear optical elements implement an 
entangling operation. This means these gates are naturally probabilistic (but 
heralded) and scales constant for each gate, hence the entire circuit scales
polynomially. Similarly, universal computation with the physical primitives shows polynomial 
scalability.  Although the physical primitives are deterministic operations, a
nonlinear coupling may realistically give a huge constant overhead for each gate.
This is simply because the physical systems/materials providing the optical 
nonlinearities are typically weak in nature (can not provide a $\pi$ nonlinear 
phase shift)\cite{boyd99}. The size of weak nonlinearity does not bring fundamental differences 
in the physical criteria, however, the subtlety here is that the polynomial scaling 
property of deterministic gates is dependent on the size of nonlinearity and hence 
the scaling property is slightly different from the one with linear optics quantum
computation (Note that the polynomial scaling with a huge overhead for each
nonlinear gate is the best scaling linear optical quantum computation can achieve.).  
This evaluation of efficiency is somewhat disturbing because the scaling is dependent
on how we count the gates.  That is if we count a set of the same nonlinear gates 
with small nonlinearity for each gate as one gate with nonlinearity large enough to 
construct CNOT gate, we have different scalabilities from one to another. 
Although the difference is only marginal being polynomial or constant and further 
the concept of scalability is a measure to distinguish the scaling from exponential 
to polynomial and/or to constant, it is still important to distinguish these 
differences in polynomial (and constant) scalability more precisely.  Physically these
differences in scalability are important to compare the potential power in the set of
physical resources. When a circuit involves probabilistic features 
in its {\it in-line} circuit, the amount of computational resources increases with 
the probability of success going to one. Roughly speaking, this give us polynomial 
scaling with a constant overhead. The case with small nonlinearity, since the gate 
is deterministic, gives a constant scaling, however each logic gate may carry a 
rather huge overhead. There is a practical problem in the scalability in these two cases, 
since both are inefficient in terms of realistic scaling even for quite large scale 
quantum computation. The fundamental question here is whether the difference between
linear optical quantum computation and nonlinear optical quantum computation is just a matter
of size of nonlinearity. We address this question next.

\begin{figure}[!htb]
\includegraphics[scale=0.4]{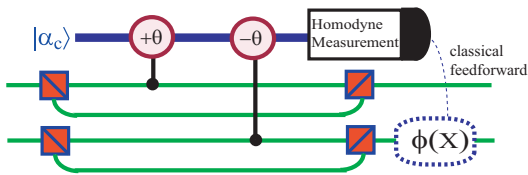}
\caption{Schematic diagram of a two qubit parity gate for qubits encoded in the polarization of
single photons.  This gate uses only linear optical elements, classical feed-forward, and weak 
cross- Kerr  nonlinearities and works as follows. Consider the four two qubit basis states 
$|HH\rangle$, $|HV\rangle$, $|VH\rangle$, $|VV\rangle$  with the probe beam in the coherent state 
$|\alpha\rangle$. After the interaction with both weak cross-Kerr nonlinearities certain 
of these basis states cause a phase shift on the probe beam. Namely the $|HV\rangle$ ($|VH\rangle$) 
basis state causes a $+\theta$ ($-\theta$) phase shift on the probe beam. An appropiate homodyne measurement 
will distinguish whether the probe beam was phase shifted or not (without determining the sign of the phase 
shift). For instance with the initial two qubit state $|HH\rangle+|HV\rangle+|VH\rangle+|VV\rangle$ our parity
gate conditions the state to either $|HH\rangle+|VV\rangle$ for the even parity result or  $e^{i \phi (X)}
|HV\rangle+e^{-i \phi (X)}|VH\rangle$ for the odd parity result. Here $\phi(X)$ is a measurement dependant phase shift 
which can be simply removed with passive optics. This shows how this 
gate can act as an entangling operation. } 
\label{primitive-gate}
\end{figure}

We might think that it should be trivial that gates have to be either deterministic 
or probabilistic (non-deterministic).  However, there is a subtlety can come in here 
when we take experimental reality into account.
Recently some logic gates using QND measurement have been shown to be 
near-deterministic\cite{Munro03,barrett04,nemoto04,munro05}. There gates are near-deterministic in a sense that they can, 
in principle, be made arbitrary close to deterministic.  Mathematically speaking, 
these gates must be counted as probabilistic, however these type 
of gates show an asymptotic failure rate behavior going to zero as a
limit. Here, we define that a gate is near-deterministic if the theoretical error 
of failure in the gate can be suppressed arbitrarily small without investing extra 
physical primitives.  Such a gate can be constructed with the physical 
primitives as Fig.(\ref{primitive-gate}). This gate works as two-qubit parity gate\cite{nemoto04,munro05}, 
which entangles the two qubits.  We call this gate a primitive gate to 
distinguish it from fundamental logic gates.  The gate is primitive 
in a sense that it is a building block of entangling gates such as the CNOT and 
Toffoli gates, but is not a physical primitive because it can be constructed by the 
combination of the physical primitives.

The state discrimination by the measurement in Fig. (\ref{primitive-gate}) on the coherent mode is not 
deterministic in a sense that the measurement result always carries theoretical 
imperfection due to the non-zero overlap in two coherent states. The failure
probability of this gate scales inversely proportional to the exponential of the distance 
between the two coherent states, and hence can be theoretically arbitrarily small and
still satisfy the condition for near-deterministic gates. Practically, the failure
probability of the gate can be controlled to be smaller than any other imperfection 
arising from physical elements in the computational circuits. This control on the 
failure probability is done by the intensity of the coherent state and hence there 
is no extra computational resources required.  Thus this gate can be considered as a
primitive gate.  Recalling the upper bounds in the success probability and No-Go
theorems in linear optics quantum computation\cite{Knill03} and quantum computation
with deterministic nonlinear gates, it is clear that the primitive gate gives us a
different scaling properties.

\begin{figure}[!htb]
\includegraphics[scale=0.3]{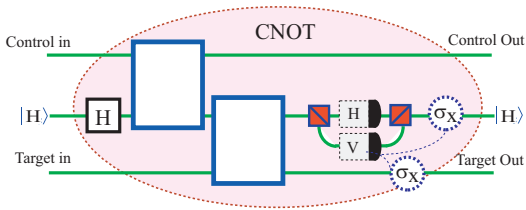}
\caption{Schematic diagram showing the construction of a CNOT gates from the 
qunat physical primitives. The gates consists of three qubits (arbitrary 
control and target qubits and a known ancilla $|H\rangle$) plus several 
key elements including the parity gates (large square box) and 
single photon QND detectors. All these elements including the single 
photon states can all be constructed from the physical primitives. A detailed 
description of how the overall gate operates can be found in  Ref \cite{nemoto04}.}
\label{cnot}
\end{figure}

It is well known that any two-qubit unitary entangling gate with local operations can 
be used  to create a CNOT gate and hence a universal set of gates\cite{Brylinski02}. However
the QND primitive gate is not a unitary gate and hence is not a fundamental logic 
gate for quantum computation in the usual sense.  For quantum computation based 
on this QND primitive gate the physical processes in a unitary gate may not be 
unitary, however the logic flow of quantum computation can remain unitary.  
To construct logic gates from the QND primitive 
gate, the second qubit in Fig (\ref{primitive-gate}) may necessarily be an 
ancilla mode for the logic gate to be carry out, however the ancilla photon 
is not destroyed in the gate and can be re-used in another gate later in the 
process and hence it is more appropriate to be considered as a part of computational
qubits. In Fig (\ref{cnot}) we illustrate how the primitive gates with the physical
primitives can construct a near-deterministic two-qubit CNOT logic gate.  
As the CNOT gate with linear optical elements is sufficient to construct a universal 
set of gates, these physical primitives can achieve universal quantum 
computation.  Thus the primitive gate is at least interchangeable with CNOT gate and
one-to-one correspondence between physical implementation to logical computational
elements. 

As we have proved that the gate set of the physical primitives is universal,  
we now need to discuss the asymptotic properties of the primitive gate. 
When the success probability of the primitive gate goes to one, 
the intensity of coherent state has to go to infinity with the given amount 
of nonlinearity in the physical primitive being constant.   
The rate of this divergence is rather slow. The error rate decreases exponentially and the amplification of the nonlinearity by intense coherent state is efficient. Considering the weak nonlinearity limit in the QND-based quantum computation, 
we notice that when the size of nonlinearity goes zero, the intensity of coherent 
state has to go infinity. The limit of zero-nonlinearity is discontinuous from the 
regime of physics primitives as the amplification of the nonlinearity breaks 
down. In contrast, gate operations in linear optics quantum computation can approach 
perfect only when the computational resource goes infinity. 

Finally we must address the issue of the generation of the single photon resources. 
The generation of these resources with the physical primitives is not deterministic nor 
near-deterministic. It is however heralded. This is a weak point in single photon optical quantum
computation.  However, single photon generation is, unlike other gates, applied
only once in the computation to generate at least $N+1$ qubits for the $N$-qubit 
computation. A simple estimation gives that allowing a certain failure probability 
for one single photon generation, $c\times N$ physical primitives are required where
$c$ is a small constant.  For instance, for the size of 10-qubit computation $c=7$
gives the order of $10^{-5}$ failure probability to produce at least $10$ single
photons, while for $N=1000$, $c=3$ gives a similar failure probability.
The factor of the constant overhead decreases even further when the computational system size
grows larger. Hence even though single-photon generation does not exhibit
the properties of near-deterministic gates, it possesses a unique scaling property 
which makes it sufficiently efficient for large scale quantum computation.

We may summarize our physical primitives for universal optical quantum 
computation:
\vspace{-2mm}
\begin{itemize}
\item Arbitrarily-intense coherent state source, \vspace{-2mm}
\item Linear optical elements, \vspace{-2mm}
\item Homodyne detection, \vspace{-2mm}
\item Non-zero cross-Kerr nonlinearity, \vspace{-2mm}
\item classical feed-forward. \vspace{-2mm}
\end{itemize}
These requirements are for universal quantum computation in optics.

{\it To conclude}, we have shown that our primitive gate with linear optical elements
is universal. This set of the gates gives the physical requirements for universal quantum 
computation in optics and these requirements can be applied to other physical systems. 
This QND-based quantum computation merges qubit(discrete)-quantum computation and 
qunat-quantum computation by having the physical primitives as the essential conditions for 
universal quantum computation. The primitive gate was shown to have a unique scaling 
property which allows the gate to construct resource efficient  logic gates such as 
the CNOT gate, that is, there is essentially no physical overhead per logic gate.  
Finally the regime of optical quantum computation given in this paper is shown to 
be distinct from all linear optics quantum computation with respect to the efficiency 
of the physical resources in the computational system.

\noindent
{\em Acknowledgments}: We will like to thank R. Beausoleil 
and T. Spiller for valuable discussions. This work was supported by a JSPS research 
grant and fellowship, and the European Project RAMBOQ.


\begin{thebibliography}{99}

\bibitem{barenco95} A. Barenco, C. H. Bennett, R. Cleve, D. P. DiVincenzo, N. H.
Margolus, P. W. Shor, T. Sleator, J. A. Smolin, and H. Weinfurter, Phys. Rev. A
{\bf 52}, 3457 (1995).

\bibitem{gk99} D. Gottesman, in {\it Proceedings of the XXII International Colloquium on 
Group Theoretical Methods in Physics}, edited by S. P. Corney et al. 
(International Press, Cambridge, MA, 1999), p. 32.
%
\bibitem{gk99b} M. A. Nielsen and I. L. Chuang, {\it Quantum Computation and 
Quantum Information}, (Cambridge University Press, Cambridge, U.K., 2000), p464.
%
\bibitem{got04} S. Aaronson and D. Gottesman, {\it Improved Simulation of Stabilizer 
Circuits}, quant-ph/0406196, 2004.
%
\bibitem{boykin00}P. O. Boykin, T. Mor, M. Pulver, V. Roychowdhury, and F. Vatan,
Information Processing Letters, 75(3), 101 (2000).
%
\bibitem{kitaev97} A. Y. Kitaev, Russian Mathematical Surveys, 52, (6), 
1191-1249 (1997).
%
\bibitem{rudolph02}  Terry Rudolph and Lov Grover, {\it A 2 rebit gate universal 
for quantum computing}, quant-ph/0210187.
%
\bibitem{Nielsen03} M. A. Nielsen, Phys. Lett. A {\bf 308}, 96 (2003).
%
\bibitem{Leung04} D. W. Leung, Int. J. Quant. Inf. {\bf 2}, 33 (2004). 
%
\bibitem{Briegel01} R. Raussendorf and H. J. Briegel, Phys. Rev. Lett. {\bf 86}, 5188 (2001).
%
\bibitem{Lloyd99} S. Lloyd and S. L. Braunstein, Phys. Rev. Lett. {\bf 82}, 1784 (1999).
%
\bibitem{KLM} E. Knill, R. Laflamme, and G. Milburn, Nature {\bf 409}, 46 (2001).
%
\bibitem{Pittman01} T.B. Pittman, B.C. Jacobs and J.D. Franson, Phys. Rev. A {\bf 64}, 062311 (2001).
%
\bibitem{Ralph02} T. C. Ralph, A. G. White, W. J. Munro, and G. J. Milburn, Phys. Rev. A {\bf 65}, 012314 (2002).
%
\bibitem{Knill02} E. Knill, Phys. Rev. A {\bf 66}, 052306 (2002).
%
\bibitem{Scheel03} Stefan Scheel, Kae Nemoto, William J. Munro, and Peter L. Knight, Phys. Rev. A {\bf 68}, 032310 (2003).  
%
\bibitem{Gottesman99} D.~Gottesman and I.~L.~Chuang, Nature {\bf 402}, 390 (1999).
%
\bibitem{Bartlett03}  Stephen D. Bartlett and William J. Munro, Phys. Rev. Lett. 90, 117901 (2003).
%
\bibitem{Bartlett02} S. D. Bartlett, B, C. Sanders, S. L. Braunstein, and Kae Nemoto, Phys. Rev. Lett. {\bf 88}, 097904 (2002).
%
\bibitem{footnote} In continuous variables 
the Hamiltonian's describing the evolution of the system are generally constructed 
from creation and destruction operators of the field. A quadratic Hamiltonians 
contains only these construction and destruction operators to second order. These 
quadratic Hamiltonian's have linear equations of motion and hence are not considered 
nonlinear in nature.
%
\bibitem{Milburn84} G. J. Milburn and D. F. Walls, Phys. Rev. A {\bf 30}, 56 (1984).
%
\bibitem{Yamamoto87} M. Kitaga, N.Imoto, and Y.Yamamoto, Phys. Rev. A 35, 5270-5273 (1987).
%
\bibitem{Munro03} W. J. Munro, Kae Nemoto, R. G. Beausoleil and T. P. Spiller, 
Phys. Rev. A 71, 033819 (2005)
%
\bibitem{boyd99} R.~W. Boyd, J. Mod. Opt. {\bf 46}, 367 (1999).
%
\bibitem{barrett04} S. D. Barrett, P. Kok, Kae Nemoto, R. G. Beausoleil, W. J. Munro and T. P. Spiller, 
Phys. Rev. A 71, 060302R (2005);.
%
\bibitem{nemoto04} Kae Nemoto and W. J. Munro, Phys. Rev. Lett {\bf 93}, 250502 (2004).
%
\bibitem{munro05} W. J. Munro, K. Nemoto and T. P. Spiller,  {\it Weak nonlinearities: 
a new route to optical quantum computation},  New J. Phys. 7, 137 (2005).
%
\bibitem{Knill03} E. Knill, Phys. Rev. A {\bf 68}, 064303 (2003).
%
\bibitem{Brylinski02} J. Brylinski and R. Brylinski, Universal quantum gates, in Mathematics of
Quantum  Computation, (edited by R. Brylinski and G. Chen), 2002

\end{thebibliography}
\end{document}